\documentclass[reqno]{article}

\usepackage[english]{babel}
\usepackage{amsmath, amsthm, amssymb, amscd, amsfonts, bm, bbm, mathrsfs}   
\usepackage[left=2cm,right=2cm,top=2cm,bottom=2cm,letterpaper]{geometry}
\usepackage{graphicx}
\usepackage[dvipsnames]{xcolor}
\usepackage{indentfirst}
\usepackage{tikz}
\usepackage{mathtools}
\usepackage[onehalfspacing]{setspace}
\usepackage{soul}
\usepackage{accents}
\usepackage{fancyhdr}
\usepackage{listings} 
\usepackage{enumerate}
\usepackage{subcaption}
\usepackage[colorlinks]{hyperref}
\usepackage{bookmark} 
\usepackage[final]{pdfpages}   
\usepackage{chngcntr}
\usepackage{fix-cm}   
\usepackage{multirow}
\usepackage{mdframed}
\usepackage[shortlabels]{enumitem}
\usepackage[authoryear]{natbib}
\usepackage{makecell}
\usepackage{arydshln}
\usepackage{xr}
\usepackage{rotating}
\usepackage{cleveref}
\usepackage{algorithm} 
\usepackage{algpseudocode}
\usepackage{booktabs}  

\hypersetup{
	breaklinks,
	linkcolor=blue,
	urlcolor=blue,
	anchorcolor=blue,
	citecolor=black
}

\theoremstyle{plain}
\newtheorem{thm}{Theorem}

\newtheorem{prop}{Proposition}

\newtheorem{rem}{Remark}

\newcommand{\ip}[1]{\left\langle #1 \right\rangle} 

\newcommand{\argmax}{\operatornamewithlimits{argmax}} 

\makeatletter
\newcommand*{\rom}[1]{\expandafter\@slowromancap\romannumeral #1@}
\newcommand{\HUGE}{\@setfontsize\Huge{40}{50}}   
\makeatother

\newcommand{\cond}{\xrightarrow{\text{d}}} 

\newcommand{\pr}{\mathsf{P}}
\newcommand{\eo}{\mathsf{E}}
\newcommand{\cov}{\mathsf{cov}}

\newcommand{\nd}{\mathsf{N}}

\newcommand{\Ga}{\Gamma} 
\newcommand{\ld}{\lambda} 

\newcommand{\HH}{\mathbb{H}} 
\newcommand{\I}{\mathbb{I}} 
\newcommand{\R}{\mathbb{R}} 

\newcommand{\mbs}{\mathbf{s}} 

\newcommand{\kll}{\mathcal{L}}	
\newcommand{\ttt}{\mathcal{T}}	

\usepackage{ulem}

\makeatletter
\newcommand{\leqnomode}{\tagsleft@true}
\newcommand{\reqnomode}{\tagsleft@false}
\makeatother

\addto\extrasenglish{%
}

\addto\extrasenglish{

}

\title{A new class of functional conditional autoregressive models}

\author{
	\small{
		Sooran Kim\footnote{Corresponding author. \\ E-mail address: \href{mailto:Sooran.Kim@nyulangone.org}{Sooran.Kim@nyulangone.org} (S. Kim)}
	} 	
}

\date{
\small{
	Division of Biostatistics, Department of Population Health, New York University, New York, NY, 10016, USA}
	}

\begin{document}
\maketitle

\begin{abstract}
We introduce a new class of conditional autoregressive models for spatially dependent functional data, 
formulated through conditional means given neighboring functional observations and characterized by a covariance operator and a spatial dependence parameter. 
Our estimation strategy consists of three components: 
(i) estimating the covariance operator using conditionally centered data, 
(ii) estimating the spatial dependence parameter by maximizing the likelihood of projected observations, 
and (iii) applying a novel profile-based approach to obtain the final estimators. 
Under an expanding lattice framework, we establish two key theoretical results. 
First, we establish the consistency of the proposed covariance estimator, 
which is not attainable using naive methods based on marginally centered data. 
Second, we prove that the spatial dependence parameter estimator is superconsistent and asymptotically normal,
where the latter property enables statistical inference for spatial dependence in functional data---a contribution that is novel in the existing literature. 
Numerical studies support the theoretical results and demonstrate the computational efficiency of our method. 
Finally, we illustrate its practical utility by analyzing weekly PM$_{2.5}$ concentration trajectories in 2019 across counties in the Midwestern United States.
\end{abstract}


\section{Introduction}

Recently, we have increasingly encountered complex data structures arising from various sources. 
One contributor is the advancement of technology enabling observation of high-resolution data, which is often categorized as functional data \cite[e.g.,][]{kokoszka2017}.
Another factor of complexity stems from spatial dependence among observational units, such as regions or areas, due to their geographical closeness---an area of study known as spatial statistics \cite[e.g.,][]{cressie2015statistics}.
Data that exhibit both functional and spatial characteristics, known as \textit{spatial functional data}, have been widely studied over the past decades \cite[cf.][]{delicado2010statistics, martinez2020recent}, as such data structures are more prevalent in practice.
For example, daily or weekly PM$_{2.5}$ concentration trajectories observed across counties in the United States can be treated as functional data, with spatial dependence deriving from the spatial proximity of counties (cf. \autoref{sec6}).
Statistical inference may be limited if such datasets are analyzed without accounting for their inherent structure—either by ignoring spatial dependence (e.g., assuming independence) or by neglecting the functional nature of the data (e.g., using simple averages of daily concentrations).

For spatial data observed on a lattice or over areal units, rather than over continuous space (i.e., geostatistical data) or as spatial point pattern data, 
dependence is typically modeled through neighborhood structures.
One of the most popular models for scalar-valued data in this setting is the conditional autoregressive (CAR) models, 
a common class of Gaussian Markov random field models
 \citep[cf.][]{gelfand2003proper, lee2011comparison, cressie2015statistics}; we refer to this as univariate CAR models hereafter.
As a special case of Markov random fields, the univariate CAR models are defined through conditional specifications; it involves specifying full conditional distributions based on neighborhood structures \cite[cf.][]{Besag1974}. Under certain conditions, the joint distribution exists in closed form, which makes it analytically tractable and thus popularly used.
However, such univariate CAR models may not be adequately applied to spatial functional data. This limitation calls for more flexible modeling frameworks beyond them.
For example, univariate CAR models can capture the spatial dependence of PM$_{2.5}$ concentrations at a single time point across the United States. However, our interest often lies in understanding the spatial dependence of daily or weekly PM$_{2.5}$ trajectories over time. This type of analysis requires modeling the data as functions, which in turn necessitates functional modeling approaches.
While there are several studies on multivariate CAR models \citep[cf.][]{mardia1988multi, gelfand2003proper, CB03}, 
there exists only one CAR model for functional data in the Bayesian framework \citep[cf.][]{zhang2016functional}.

The primary goal of this paper is to propose a new class of functional conditional autoregressive (FCAR) models
that extend the classical univariate CAR models into the functional data framework.
Our novel models define spatial dependence through conditional means given neighboring functional observations,
maintaining a structure that conceptually parallels that of traditional CAR models.
We derive the joint distribution of FCAR observations in analogy with finite-dimensional counterparts \citep[cf.][]{mardia1988multi, cressie2015statistics},
treating the collection of the functional observations as a multivariate random function valued in the $n$-dimensional Euclidean space.

The proposed models are clearly contrasted with the only existing FCAR models introduced by \citet{zhang2016functional}.
Upon applying basis expansion and obtaining coefficients of basis functions, they apply the univariate CAR model to the finite-dimensional vectors of the coefficients. However, this approach may be restricted to incorporating infinite-dimensionality, residing in functional data. 
While useful, the limitations of the strategies that first truncate the original functional data within the model and then apply an existing finite-dimensional method have been pointed out in other functional data literature; they may introduce model bias by neglecting the contribution of higher-order components beyond the truncation level \citep[cf.][]{DPZ12}.
We address this issue by imposing conditional model specifications directly on spatial functional data without any dimension reduction at the modeling stage.
Furthermore, our proposed method is developed within a frequentist framework, whereas \citet{zhang2016functional} adopt a Bayesian approach. 
As a consequence, our frequentist framework avoids Bayesian sampling and facilitates a detailed asymptotic analysis for inferential testing.
Further comparisons are provided in the Section~S5 in the supplement.

The proposed FCAR models include two key parameters:
a covariance operator that incorporates variability in each functional datum and a spatial dependence parameter that quantifies spatial dependence structure among functional observations.
For their estimation, we introduce a novel profile-based estimation approach. 
As the spatial structure is modeled through the conditional means of spatial functional data, the covariance operator is estimated by the empirical covariance of the \textit{conditionally} centered data (i.e., data centered by their conditional means).
This new strategy provides a substantial difference from other covariance estimators under independence, e.g., the empirical covariance of marginally centered data. 
We then estimate the spatial dependence parameter by the maximum likelihood estimator based on the log-likelihood of the data projected onto the eigenfunctions of the covariance operator.
This approach parallels existing functional data methods that apply finite-dimensional techniques to the projections \citep[cf.][]{FHKS14, KSM16}.
Nevertheless, 
our method is distinct in that the eigenfunctions used for projection are derived from the empirical covariance operator of the conditionally centered data, rather than from marginally centered data.
The final estimators are obtained by alternating between these two estimation steps until convergence. 
Overall procedure can be quite efficient compared to the finite-dimensional cases,
as it does not need (i) Bayesian tools, (ii) computation of inverse matrices, and (iii) optimization with high-dimensional variables.
Our numerical studies reveal that the algorithm converges very quickly,
despite its iterative nature.
This leads to highly efficient computations in practice.

We further develop the asymptotic theory for the Gaussian FCAR model with scalar spatial dependence in a lattice/areal setting. 
Specifically, the asymptotic properties of our estimators are established in an \textit{expanding} lattice asymptotic context---the lattice of spatial locations grows without bound. 
As the truncation level appropriately diverges depending on the sample size $n$ \cite[cf.][]{FHKS14, KSM16}, 
our covariance estimator still shows root-$n$ consistency.
One of the key novelties of our approach lies in the spatial dependence estimator, which is the superconsistent---converging at a rate faster than root-$n$.
Moreover, this estimator is asymptotically normal upon suitable scaling.
By investigating the asymptotic behaviors of both estimators, this article provides a valuable advancement to the incorporation of spatial dependence in functional data analysis.
In particular, based on these results, we propose a useful hypothesis testing to assess the presence of spatial dependence in functional data.
To the best of our knowledge, this is the first attempt to develop a hypothesis testing for the global spatial dependence in functional data. 
In spatial functional data analysis, this test will serve an analogous role to Moran’s I test,
which has been widely used for testing spatial dependence in univariate spatial statistics \citep[cf.][]{moran1950notes}. Consequently, it fills an important methodological gap.

Our proposal also makes a significant contribution to analyzing spatially dependent functional data, which arises in numerous fields such as environmental science and public health \citep[cf.][]{liang2021modeling, tang2022functional}. 
In such problems, neglecting either the temporal or spatial dependence can result in substantial information loss, leading to reduced interpretability and potentially misleading conclusions.
To illustrate the practical utility of our approach, we apply our method to weekly averaged PM$_{2.5}$ concentration levels recorded in 2019 across counties in the Midwestern United States. Given the temporal trends and spatial correlation among neighboring counties, our analysis illustrates that ignoring either temporal or spatial structure may obscure important dependence patterns. By modeling both functional trajectories and spatial dependence, the proposed method can provide a more informative assessment of the data.


\section{Functioanl conditioanl autoregressive models}\label{sec2}

In this section, we propose a new class of CAR models for spatial functional data.
To describe, let $\HH=\kll_2(\ttt) \equiv \{f : \ttt \rightarrow \R| \int_{\ttt} f^2(t) dt < \infty \}$ be the infinite-dimensional separable Hilbert space of all square integrable functions on the compact interval $\ttt$ in the real line, 
equipped with inner product $\ip{f_1, f_2}_{\HH} \equiv \int_{\ttt} f_1(t)f_2(t)dt$ for $f_1, f_2 \in \HH$. 
The functional responses are assumed to take values in $\HH$ and observed at $n$ spatial locations $\{\mbs_i\}_{i=1}^n$, where each individual response is denoted by $Y(\mbs_i) = \{Y(\mbs_i)(t): t \in \ttt\}$ for $i=1,\cdots, n$.

Similar to the univariate CAR model by \cite{Besag1974} and multivariate counterpart by \cite{mardia1988multi},
a FCAR model for $\{Y(\bm{s}_i)\}_{i=1}^n$ is defined through conditional distributions of each functional observation $Y(\bm{s}_i)$ given the rest $Y(\bm{s}_{-i}) \equiv \{ Y(\bm{s}_j): j \neq i \}$ of the observations 
as
\begin{align}\label{model:full conditional}
	Y(\mbs_i) | Y(\mbs_{-i}) 
	\sim \mathsf{Gaussian} (\mu_i, \Gamma_i), 
\end{align}
where the conditional mean $\mu_i \equiv \eo[Y(\mbs_i) | Y(\mbs_{-i})]$ is given by
\begin{align}\label{model:conditioanl mean}
	\mu_i &= \alpha_i + \sum_{j\neq i} \eta_{ij} (Y(\mbs_j)-\alpha_{j}).
\end{align}
Here, $\alpha_i \in \HH$ is shown to be the marginal mean of $Y(\bm{s}_i)$
and $\Gamma_i:\HH\to\HH$ is a self-adjoint, non-negative definite, and Hilbert--Schmidt operator on $\HH$ that plays a role of the conditional covariance of $Y(\bm{s}_i)$ \cite[cf.][]{HE15}. 
The parameters $\eta_{ij}$ and $\Gamma_i$ are subject to the following conditions:
$\eta_{ij}\Gamma_j = \eta_{ji} \Gamma_i$ to ensure symmetry, 
$\eta_{ii}=0$ for $i=1,\cdots, n$ to avoid self-dependence, and
$\eta_{ij}=0$ unless $\mbs_j \in N_i$, where $N_i = \{\mbs_j:\mbs_j \text{ is a neighbors of }\mbs_i \}$, so that dependence is restricted to neighboring locations.
Thus, our proposed FCAR models are characterized by conditional mean elements and conditional covariance operators, incorporating spatial neighborhood information, which introduces a novel modeling framework for spatial functional data.

We extend the construction of the joint distribution of the FCAR observations.
To describe the joint distribution, we further define another Hilbert space $\HH(\R^n)=\kll_2(\ttt, \R^n) \equiv \{\tilde{f} : \ttt \rightarrow \R^n | \int_{\ttt} \|\tilde{f}(t)\|_{n}^2 dt < \infty \}$, as the space of $\R^n$-valued square integrable functions on $\ttt$, where $\tilde{f}(t) = [f_1(t), \cdots, f_n(t)]^\top$ for $t \in \ttt$ with $f_i \in \HH$, $i=1,\dots,n$. 		
The inner product in $\HH(\R^n)$ is defined analogously as  
$\langle {\tilde{f}, \tilde{g}}\rangle \equiv \int_{\ttt} \sum_{i=1}^n f_i(t)g_i(t)dt=\sum_{i=1}^n \ip{f_i, g_i}_{\HH}$ for $\tilde{f}, \tilde{g} \in \HH(\R^n)$.
Within this framework, the $\HH(\R^n)$-valued random variable 
$\tilde{Y}(\omega): \ttt \rightarrow \R^n$, 
\begin{align*}
	(\tilde{Y}(\omega))(t) = [Y(\mbs_1)(\omega)(t), \cdots, Y(\mbs_n)(\omega)(t)]^\top, \quad \forall t \in \ttt.
\end{align*}
represents the stacked functional observations across spatial locations.
\autoref{prop:joint} provides the joint distribution of this stacked functional observations $\tilde{Y}$.

\begin{prop}
	\label{prop:joint}
	With the matrix of spatial dependence parameters $\bm{H}=[\eta_{ij}]_{1\leq i,j \leq n}$, 
	let $\Delta_{j} = \sum_{i=1}^n \left\{(\bm{I}_n - \bm{H})^{-1}\right\}_{ij} \Gamma_j:\HH \to \HH$, and
	define the operator 
	$\tilde{\Delta} : \HH(\R^n) \to \HH(\R^n)$ by 
	$(\tilde{\Delta} \tilde{f})(t) = [({\Delta}_{1} {f}_1)(t), \cdots, ({\Delta}_{n} f_n)(t)]^\top$.
	Then, the joint distribution of $\tilde{Y}$ is derived as
	\begin{align*}
		\tilde{Y} \sim \mathsf{Gaussian}(\tilde{\alpha}, \tilde{\Delta}),
	\end{align*}
	where $\tilde{\alpha}(t)=[\alpha_1(t), \cdots, \alpha_n(t)]^\top$.
\end{prop}
Details of the joint distribution derivation are provided in Section~S2 of the supplement.
This joint distribution is a generalization of univariate \citep[cf.][Equation (6.6.4)]{cressie2015statistics} and multivariate counterparts \citep[cf.][Theorem~2.1]{mardia1988multi}.
Unlike these finite-dimensional CAR models, the covariance in our model is represented by a bounded linear operator on the space of vector-valued functions, rather than a finite-dimensional covariance matrix. 
It preserves the core intuition of classical CAR models while accommodating the infinite-dimensional nature of functional data.

We now present an example of FCAR models for spatial functional data.
	Suppose that the model components are homogeneous across locations:
	$\alpha_i = \alpha$ and $\Gamma_i = \Gamma$ for all $i$.
	Moreover, we specify constant spatial dependence $\eta_{ij} = \eta \I(i \sim j)$, where $\I(i \sim j)=1$ if locations $i$ and $j$ are neighbors and 0 otherwise.
	In this case, the matrix $\bm{H} = [\eta_{ij}]_{1 \leq i,j \leq n}$ reduces to $\eta \bm{W}$, 
	where
	$\bm{W} = [w_{ij}]_{1 \leq i, j \leq n} $ is a symmetric neighborhood matrix defined by
	\begin{align}\label{def:w}
		w_{ij} = \begin{cases}
			1 	&	\text{if $\mbs_{i}$ and $\mbs_{j}$ are neighbors}
			\\ 0 & \text{otherwise}
		\end{cases}.
	\end{align}
	That is,
	\begin{align}\label{general_model}
		Y(\mbs_i) | Y(\mbs_{-i}) \sim \mathsf{Gaussian} (\mu_i, \Gamma), 
	\end{align}
	where the conditional mean is given by
	\begin{align*}
		\mu_i &= \alpha + \eta \sum_{j \in N_i}  (Y(\mbs_j)-\alpha).
	\end{align*}


An analogous formulation is widely used in the univariate setting; see \cite{cressie2015statistics}. 
We adopt this model throughout the estimation and theoretical developments in Sections~\ref{sec3}–\ref{sec4}.


\section{Estimation}\label{sec3}

We introduce our estimation procedure for $\Gamma$ and $\eta$, 
which is computationally efficient and yields accurate estimators. 
The method employs an alternating algorithm;
the parameters $\Gamma$ and $\eta$ of main interest are alternatively updated given the other until convergence. 
The procedure is grounded by the following two key ideas.
First, to estimate $\Gamma$, we use conditionally centered responses, 
as they possess simpler dependence (cf.~Lemma~S1 in the supplement).
Second, utilizing the Gaussian nature of CAR models, 
the projections onto the eigenfunctions of $\Gamma$ are independent.
This independence enables the estimation of $\eta$ through the likelihood derived from $p$ independent $n$-dimensional CAR random vectors. 
The details are described as follows. 

Given $\alpha \in \HH$ and $\eta \in \R$, 
we define conditionally centered responses $\{Z(\bm{s}_i)\}_{i=1}^n$ as
\begin{align}\label{Z}
	Z(\bm{s}_i)
	= Z(\bm{s}_i; \alpha, \eta)
	& \equiv \{ Y(\bm{s}_i) - \alpha \} - \eta \sum_{j \in N_i} \{Y(\bm{s}_j) - \alpha \}
\end{align}
so that $Z(\bm{s}_i) \sim \mathsf{Gaussian}(0,\Gamma)$. 
It is worth noting that, as shown in Lemma~S1 in the supplement, the conditionally centered data $\{Z(\bm{s}_i)\}_{i=1}^n$ exhibit an simpler dependence structure than the original data $\{Y(\bm{s}_i)\}_{i=1}^n$.
Namely, it holds that 
\begin{align*}
	\cov[Z(\bm{s}_i), Z(\bm{s}_j)] = -\eta \I(i \sim j)\Gamma,
\end{align*}
meaning that the conditionally centered data are correlated only when they are neighbors.
Inspired these distributional results for $\{Z(\bm{s}_i)\}_{i=1}^n$,
we suggest estimating $\Gamma$ 
by the empirical covariance operator of $\{Z(\bm{s}_i)\}_{i=1}^n$:
\begin{align} \label{eqGaHatCondCent}
	\hat{\Gamma}(\alpha, \eta)
	\equiv {1\over n} \sum_{i=1}^n (Z(\mbs_i) - \bar{Z}) \otimes (Z(\mbs_i) - \bar{Z}),
\end{align}
where
$\bar{Z}=\bar{Z}(\alpha, \eta) \equiv n^{-1} \sum_{i=1}^n Z(\mbs_i)$.
From this observation, we may expect that the covariance estimator $\hat{\Gamma}(\alpha, \eta)$ in \eqref{eqGaHatCondCent} would behave quite regularly and become accurate (cf.~Proposition~S1 in the supplement).
In contrast, the naive empirical covariance estimator 
\begin{align} \label{eqGaHatMargCent}
	\hat{\Gamma}_{\mathrm{naive}}
	\equiv {1\over n} \sum_{i=1}^n (Y(\mbs_i) - \bar{Y}) \otimes (Y(\mbs_i) - \bar{Y})
\end{align}
constructed by $\{Y(\bm{s}_i)\}_{i=1}^n$ may perform poorly under strong spatial dependence.
See \autoref{sec5} for a substantial numerical difference between these two covariance estimators.

Next, we suppose $\alpha \in \HH$ and $\Gamma:\HH\to\HH$ are given,
where $\Gamma$ admits a spectral decomposition as
\begin{align*}
	\Gamma = \sum_{j=1}^\infty \ld_j (\phi_j \otimes \phi_j),
\end{align*}
where
$\lambda_1 \geq \lambda_2 \geq \cdots 0$ are the eigenvalues and 
$\{\phi_j\}_{j=1}^\infty$ are the orthonormal eigenfunctions of $\Gamma$.
Here, $x \otimes y:\HH\to\HH$ represents the tensor product between $x,y \in \HH$, the rank one operator defined as  $(x \otimes y)(z) = \langle z, x \rangle_{\HH} y$ for $z \in \HH$.  
The projections $\{ \langle Y(\bm{s}_i), \phi_j \rangle_{\HH} \}_{i=1}^n$ onto each eigenfunction then form a univariate CAR model and are determined as
\begin{align*}
	\bm{Y}^{\mathrm{proj}}_j  
	= \bm{Y}^{\mathrm{proj}}_j (\Ga)
	\equiv 
	\begin{bmatrix}
		\langle Y(\mbs_1), \phi_j \rangle_{\HH}
		\\ \vdots
		\\ \langle Y(\mbs_n), \phi_j \rangle_{\HH}
	\end{bmatrix}		
	\sim
	\nd_n\left(
	\bm{0},			
	\ld_j (\bm{I}_n - \eta \bm{W})^{-1} 
	\right)
\end{align*}
\cite[cf.][]{cressie2015statistics}.
Since the random vectors $\{ \bm{Y}^{\mathrm{proj}}_j \}_{j=1}^\infty$ are independent due to the orthogonality of $\{\phi_j\}_{j=1}^\infty$ and normality,
the log-likelihood based on the first $p$ projections $\{ \bm{Y}^{\mathrm{proj}}_j \}_{j=1}^p$ is given by
\begin{align}
	l_n(\eta)
	= l_n(\eta; \alpha,\Gamma)
	\equiv -{1\over 2n} \sum_{j=1}^p \left\{
	\ld_j^{-1} \bm{Y}^{\mathrm{proj} \top}_j \bm{\Sigma}^{-1} \bm{Y}^\mathrm{proj}_j  + \log(\det(\ld_j \bm{\Sigma}))
	\right\}, \label{eqProjLik}
\end{align}
where $\bm{\Sigma} = \bm{\Sigma}(\eta) \equiv (\bm{I}_n - \eta \bm{W})^{-1}$.
In practice, the truncation level $p$ can be chosen by the fraction of variance explained (FVE), with a threshold such as 0.95.
For estimating the spatial parameter $\eta$, we propose the maximum likelihood estimator defined by
\begin{align}
	\hat{\eta}(\alpha, \Gamma)
	\equiv \argmax_{\eta \in \R} l_n(\eta; \alpha,\Gamma). \label{eqEtaHatProjLik}
\end{align}
We emphasize that the above problem is a univariate optimization problem, which can be solved efficiently even when the sample size $n$ is large.

Finally, we introduce our profile-based estimation approach, 
wherein the covariance operator $\Gamma$ and the spatial dependence parameter $\eta$ are iteratively updated by alternating between their respective estimation steps until convergence.
Specifically, upon setting $\hat{\alpha} = n^{-1} \sum_{i=1}^n Y(\bm{s}_i)$ being the empirical average,
the updated estimators are
\begin{align*}
	\hat{\Gamma}^{(t+1)} = \hat{\Gamma}(\hat{\alpha}, \hat{\eta}^{(t)}),
	\quad
	\hat{\eta}^{(t+1)} = \hat{\eta}(\hat{\alpha}, \hat{\Gamma}^{(t+1)})
\end{align*}
given the current estimators $\hat{\eta}^{(t)}$ and $\hat{\Gamma}^{(t)}$.
We conduct this alternating procedure until both estimators converge. 
The overall estimation procedure is summarized in \autoref{algorithm:est2}.

\begin{algorithm}[htbp]
	\caption{Estimation algorithm based on the proposed profile-based approach}
	\label{algorithm:est2}
	\begin{algorithmic}[1] 
		
		\State Estimate $\alpha$ by the sample mean $\hat{\alpha} = n^{-1} \sum_{i=1}^n Y(\mathbf{s}_i).$
		
		\Statex
		\Statex \textit{Initialization}
		\State Set initial values for $\hat{\eta}^{(0)}$ and $\hat{\Gamma}^{(0)}$.
		
		\Statex
		\Statex \textit{Begin iteration}
		\State Compute the conditionally centered data $\{\hat{Z}(\mathbf{s}_i;\hat{\alpha}, \hat{\eta}^{(t)})\}_{i=1}^n$ in \eqref{Z}. 
		
		\State Compute the sample covariance of $\{\hat{Z}(\mathbf{s}_i;\hat{\alpha}, \hat{\eta}^{(t)}))\}_{i=1}^n$ as $\hat{\Gamma}^{(t+1)} = \hat{\Gamma}(\hat{\alpha},\hat{\eta}^{(t)})$ in \eqref{eqGaHatCondCent}.
		
		\State Perform spectral decomposition of the sample covariance $\hat{\Gamma}^{(t+1)}$ to obtain estimated eigenvalues $\{\hat{\lambda}_j\}_{j=1}^{p}$ and eigenfunctions $\{\hat{\phi}_j\}_{j=1}^{p}$. The truncation level $p$ is determined by the fraction of variance explained (FVE) with a threshold of $0.95$.
		
		\State Obtain the estimate $\hat{\eta}^{(t+1)} = \hat{\eta}(\hat{\alpha},\hat{\Gamma}^{(t+1)})$ as in \eqref{eqEtaHatProjLik} by maximizing the log-likelihood $l_n(\cdot; \hat{\alpha}, \hat{\Gamma}^{(t+1)})$ given in \eqref{eqProjLik}.
		
		\Statex
		\Statex \textit{End iteration if $|\hat{\eta}^{(t+1)} - \hat{\eta}^{(t)}| \leq \tau_1$ and $\| \hat{\Gamma}^{(t+1)} - \hat{\Gamma}^{(t)} \|_{\mathrm{HS}} \leq \tau_2$ for some small threshold values $\tau_1,\tau_2 > 0$.}
		
	\end{algorithmic}
\end{algorithm}

\begin{rem} 
	The procedure outlined in \autoref{algorithm:est2} could extend naturally to inference in multivariate CAR models,
	where estimating numerous covariance parameters typically entails high computational costs.
	From this perspective, \autoref{algorithm:est2} may also be useful for classical CAR models, since it
	(i) avoids Monte Carlo approximations for Bayesian inference \citep[cf.][]{sain2011spatial}, 
	and (ii) does not require large-scale optimization over many parameters \citep[cf.][]{cressie2015statistics}. 
\end{rem}

\begin{rem}
	The proposed framework can be extended to regression with scalar covariates, corresponding to a function-on-scalar regression setting.
	Specifically, suppose that the marginal mean of $Y(\mbs_i)$ in \eqref{model:conditioanl mean} is given by 
	$$\alpha_i(t) = \bm{X}_i^\top \tilde{\beta}(t),$$ 
	where $\bm{X}_i = [X_{i1}, \cdots, X_{iq}]^\top$ denotes a vector of scalar covariates and 
	$\tilde{\beta}(t) = [\beta_1(t), \cdots, \beta_q(t)]^\top$, with $\beta_j \in \HH$ for all $j=1,\cdots, q$.
	Let $\bm{X}=[\bm{X}_1, \cdots, \bm{X}_n]^\top$, and 
	let $\tilde{Y}(t)=[Y(\mbs_1)(t),\cdots, Y(\mbs_n)(t)]^\top$ be the vector of functional responses evaluated at time $t$.
	Then, the least squares estimator of $\tilde{\beta}(t)$ is given by 
	\begin{align*}
		\hat{\tilde{\beta}}(t) = (\bm{X}^\top \bm{X})^{-1}\bm{X}^\top \tilde{Y}(t),
	\end{align*}
	provided that $\bm{X}^\top \bm{X}$ is invertible \citep[cf.][]{zhang2007statistical, zhang2011statistical}.
	Accordingly, the marginal mean $\alpha_i$ is estimated as
	$\hat{\alpha}_i(t) = \bm{X}_i^\top \hat{\tilde{\beta}}(t)$,
	replacing the sample mean used in \autoref{algorithm:est2}.

	This regression formulation is intended only to indicate how the marginal mean estimation step may be modified when scalar covariates are available.
	In the following section, the theoretical results are established under the common-mean case, with the main focus on estimation and inference for the spatial dependence parameter $\eta$ and covariance operator $\Gamma$.
	A full asymptotic theory for the more general setting is beyond the scope of the present paper.
	
\end{rem}

\section{Asymptotic theory}\label{sec4}

In spatial statistics with discrete spatial indices, there have been three common contexts for asymptotic theory: 
the \textit{repeating}, \textit{infill}, and \textit{expanding} asymptotic contexts \citep[cf.][]{varin2011overview, cressie2015statistics}. 
The repeating asymptotic context indicates increasing sample size through independent realizations of a fixed grid,
so that asymptotic arguments rely on an increasing number of i.i.d realizations observed at the same set of spatial locations.
The infill (or fixed-domain) asymptotic context assumes that observations become increasingly dense in a fixed and bounded grid,
typically by refining the lattice spacing while keeping the spatial domain unchanged.
The expanding lattice (or increasing-domain) context assumes that the lattice of spatial locations grows without bound, 
while the spacing between neighboring sites remains bounded away from zero.
In particular, both the infill and expanding lattice contexts involve asymptotics based on a single observed realization with an increasing number of spatial locations, in contrast to the repeating lattice context.
The infill asymptotic context is especially common in geostatistics and spatial econometrics,
where interest typically lies in interpolation or prediction over a fixed region \citep[cf.][]{stein1999interpolation, lee2004asymptotic}. 
On the other hand, many spatial datasets—especially those defined on county-level or nation-level areas—are more naturally interpreted under an expanding lattice asymptotic context
\citep[cf.][]{gaetan2010spatial, cressie2015statistics}.
Thus, this paper focuses on the expanding lattice context.

We here present main results focusing on the asymptotic properties of $\hat{\Ga}(\alpha, \eta)$ in \eqref{eqGaHatCondCent} and $\hat{\eta}(\alpha, \Gamma)$ in \eqref{eqEtaHatProjLik}, which theoretically support \autoref{algorithm:est2} for the Gaussian FCAR model with scalar spatial dependence under the expanding lattice/areal setting.
Since our primary interest is the estimation of dependence structures represented by $\Ga$ and $\eta$, 
the subsequent discussions will proceed assuming $\alpha = 0$.
To describe, we list some technical assumptions below.

\begin{enumerate}[label=(C\arabic*), ref=(C\arabic*)]
	 \setcounter{enumi}{-1}
	\item The truncation level $p=p(n)$ grows more slowly than the sample size $n$, i.e., $p^{-1} + n^{-1}p\to 0$ as $n\to\infty$; \label{C0}
\end{enumerate}

\begin{enumerate}[label=(C\arabic*), ref=(C\arabic*)]
	\item The fourth moment of the response is finite in the limit in the sense that $\limsup_n \max_{1\leq i \leq n} \eo \left\| Y(\mbs_i)\right\|_{\HH}^4 < \infty$ ; \label{A2}	
	
	\item For the sequence of eigengaps defined as 
	$\delta_1=\lambda_1-\lambda_2$ and $\delta_j = \min\{\lambda_{j}-\lambda_{j+1}, \lambda_{j-1}-\lambda_{j} \}$ for $j \geq 2$,
	$n^{-1/2} \sum_{j=1}^p  \delta_j^{-2}  \rightarrow 0$ as $n \rightarrow \infty$.	\label{A3}
\end{enumerate}

Assumption~\ref{C0} is related to the growth rates of $n$ and $p$. Throughout this section, we assume this assumption implicitly and do not restate it in individual theorems.
Assumption~\ref{A2} aligns with a standard moment condition that can be frequently observed in functional data analysis.
For instance, under independence, Assumption~\ref{A2} is equivalent to the typical finite fourth moment condition \citep[cf.][]{hall2007methodology, horvath2012inference, lin2025hypothesis}.
Assumption~\ref{A3} describes the decay rate of eigengaps $\{\delta_j\}$. 
Since the eigenvalues $\{\lambda_j\}$ decreases to zero as $j\rightarrow \infty$, 
consequently the eigengaps $\{\delta_j\}$ typically shrink for higher-order components. 
Assumption~\ref{A3} regulates the rate of this decay relative to the sample size $n$ and the number of retained components $p$.
Specifically, this assumption ensures that the cumulative effect of small eigengaps remains asymptotically negligible, so that eigenvalue and eigenfunction estimation errors do not accumulate faster than the information provided by the sample.
It is analogous to Assumption~(A6) in \cite{yeon2023bootstrap},
where the independent scenario requires $n^{-1} \sum_{j=1}^p  \delta_j^{-2}  \rightarrow 0$ as $n \rightarrow \infty$. 
For further discussion of practical implications of such a decay condition, see \cite{FHKS14}, which adopts an eigengap condition requiring a more slowly growing rate of truncation level even in an independent setting.

In addition, we impose some conditions regarding the neighborhood structure from $\{N_i\}_{i=1}^n$ that determines the spatial dependence in FCAR models. 
To explain, we define the Frobenius norm $\|\cdot \|_{\mathrm{F}}$ on the space of matrices as 
$\|\bm{A}\|_{\mathrm{F}} = (\sum_{i=1}^n \sum_{j=1}^m a_{ij}^2)^{1/2}$ for a matrix $\bm{A}=[a_{ij}]_{1\leq i \leq n, 1 \leq j \leq m}$.

\begin{enumerate}[label=(D\arabic*), ref=(D\arabic*)]
	\item
	The number of neighbors is asymptotically bounded in the sense that
	$\limsup_n \max_{1\leq i \leq n} |N_i| < \infty$, where $|N_i|$ represents the number of neighbors at location $\mbs_i$.	 \label{A1}

	\item
	The CAR covariance matrix is asymptotically bounded, i.e., 
		$\|(\bm{I}_n - \eta \bm{W})^{-1} \|_{\mathrm{F}} = O(1)$.		\label{A4}
\end{enumerate}

Assumption~\ref{A1} states that each spatial location $\mbs_i$ has a bounded number of neighbors. 
It captures local spatial dependence: observations may be correlated within a small neighborhood, but long-range interactions are excluded.
This prevents dependence from accumulating as the sample size increases and ensures that no single location is influenced by an unbounded portion of the domain.
This assumption is natural in many spatial settings, particularly for lattice or areal data, where neighborhood structures such as four- or eight-nearest neighbors are commonly used \citep[cf.][]{Besag1974, hoshino2024functional}.
Under such constructions, the number of neighbors remains fixed as the lattice expands, making the assumption compatible with expanding lattice asymptotic context. 
Assumption~\ref{A4} imposes a bound on the global spatial dependence in the Frobenius norm sense.
In standard spatial statistics, $(\bm{I}_n - \eta \bm{W})^{-1}$ often corresponds to the covariance matrix of the joint distribution from the univariate CAR models \citep[cf.][]{Besag1974}.
Thus, this assumption ensures that the covariance matrix remains bounded even as the number of spatial locations $n$ increases. 
In replacing of $\alpha$-mixing condition, it implies a sufficiently fast decay in spatial dependence, thereby limiting the long-range dependence.


\autoref{consistency:Gamma_Zhat} next establishes the consistency of the covariance estimator from \eqref{eqGaHatCondCent} based on conditionally centered data. 
To explain, we write $\|\cdot\|_{\mathrm{HS}}$ to denote the Hilbert--Schmidt (HS) norm for bounded linear operators on $\HH$.
Namely, for a bounded linear operator $T:\HH\to\HH$ such that $\sum_{j=1}^\infty \|T \psi_j\|_{\HH}^2<\infty$, 
we define
$\|T\|_{\mathrm{HS}} = (\sum_{j=1}^\infty \|T \psi_j\|_{\HH}^2)^{1/2}$,
where $\{ \psi_j \}_{j=1}^\infty$ is any complete orthonormal system for $\HH$.

\begin{thm}\label{consistency:Gamma_Zhat}
	Suppose 
	that there exists a consistent estimator $\tilde{\eta}$ for $\eta$, i.e., $\tilde{\eta} \xrightarrow{p} \eta$ as $n\to\infty$, and
	that Assumptions~\ref{A2} and \ref{A1} hold.
	Then, the covariance estimator $\hat{\Ga} = \hat{\Ga}(0, \tilde{\eta})$ from \eqref{eqGaHatCondCent} is consistent for $\Ga$ with $\sqrt{n}$-rate. More precisely, as $n\to\infty$, we have
	\begin{align*}
		\eo[ \| \hat{\Ga} - \Ga \|_{\mathrm{HS}}^2 ] = O(n^{-1}). 
	\end{align*}
\end{thm}

\autoref{consistency:Gamma_Zhat} is a consequence of the simpler dependence structure of the conditionally centered data $\{Z(\bm{s}_i)\}_{i=1}^n$ given in \eqref{Z}.
More specifically, Lemma~S1 in the supplement states that $Z(\bm{s}_i)$ and $Z(\bm{s}_{i'})$ are correlated only when the corresponding locations $\bm{s}_i$ and $\bm{s}_{i'}$ are neighbors. 
Subsequently, it can be shown that the expansion of the mean square HS error is asymptotically bounded by $n^{-1}\max_{1 \leq i \leq n} |N_i| $.
This finding is important 
as it motivated us to develop the fast algorithm in \autoref{algorithm:est2}. 

In \autoref{thm:eta} below, we provide a central limit theorem for the spatial dependence estimator $\hat{\eta}$ from \eqref{eqEtaHatProjLik}, after deriving its consistency.
This result may be of particular interest in the spatial context.

\begin{thm}\label{thm:eta}
	Suppose 
	that there exists a $\sqrt{n}$-consistent estimator $\tilde{\Ga}$ of $\Ga$ such that $\|\tilde{\Ga} - \Ga\|_{\mathrm{HS}}^2 = O_p(n^{-1})$
	and that Assumptions~\ref{A2}-\ref{A3} and \ref{A1} hold.
	Then, we have the following.
	\begin{enumerate}[label=(\alph*)]
		\item 
		The spatial dependence estimator $\hat{\eta} = \hat{\eta}(0,\tilde{\Ga})$ from \eqref{eqEtaHatProjLik} is consistent for $\eta$, i.e., as $n\to\infty$, 
		\begin{align*}
			\hat{\eta} \xrightarrow{p} \eta.
		\end{align*}
		
		\item 
		Furthermore, suppose that Assumption~\ref{A4} hold along with $n^{-1}p^3 \rightarrow 0$ as $n\rightarrow \infty$.
		Then, the spatial dependence estimator $\hat{\eta} = \hat{\eta}(0,\tilde{\Ga})$ from \eqref{eqEtaHatProjLik} is asymptotically normal with mean $\eta$ and variance $\{ - n \hat{l}_n''(\hat{\eta}) \}^{-1}$,
		i.e., as $n\to\infty$, 
		\begin{align*}
			\sqrt{n} \left(-\hat{l}_n''(\hat{\eta}) \right)^{1/2} (\hat{\eta}-\eta) \cond \nd (0, 1),
		\end{align*}
		where $\hat{l}_n''(\hat{\eta}) \equiv l_n''(\hat{\eta};0, \tilde{\Ga})$ is the second derivative of the log-likelihood $\hat{l}_n(\hat{\eta}) \equiv l_n(\hat{\eta};0,\tilde{\Ga})$ from \eqref{eqProjLik}.
		
	\end{enumerate}
\end{thm}

We discuss several theoretical observations from \autoref{thm:eta}.
First, a key feature of the log-likelihood in \eqref{eqProjLik} is the presence of a weighted sum of $p$ quadratic forms, each involving $n$-dimensional vectors of projections.
This structure significantly makes the proofs more challenging.
Each quadratic form is defined using the  covariance matrix of the univariate CAR model, $\bm{\Sigma} = \bm{\Sigma}(\eta) \equiv (\bm{I}_n - \eta \bm{W})^{-1}$, which reflects the underlying spatial dependence.
This is especially contrasted to the cases of independent functional data,
where chi-squared tests have been developed with the test statistic being a weighted sum of squared projections \citep[cf.][]{cardot2003testing, horvath2009two, FHKS14, KSM16}.
In our context, however, due to spatial dependence, it is natural to expect more intricate objects, such as the sums of \textit{quadratic forms} involving $\bm{\Sigma}$, rather than simple sums of squares. 

Another notable feature is that the estimator $\hat{\eta}$ is the superconsistent, which means it converges at a rate faster than the usual $\sqrt{n}$.
Specifically, $\hat{\eta}$ is $\sqrt{np}$-consistent from Equation~(S51) in the supplement. 
This phenomenon arises due to a sum of $p$ quadratic forms in the log-likelihood \eqref{eqProjLik}.
In contrast, such superconsistent is not observed in finite-dimensional CAR models or in classical functional linear regression under independence. 
Instances of superconsistency have been studied in other contexts. 
For example, involving unit roots or cointegration,
see \citet[Chapter~17]{Hamil94} for time series and \cite{YJL12, BFF12} for spatial context. 
In classical functional data analysis, \cite{Poss20} discusses superconsistency in the context of functional regression impact points.
There results arise in settings that are fundamentally different from ours.

Based on \autoref{thm:eta}, we can construct confidence intervals for the spatial dependence parameter $\eta$.
An approximate $(1-\kappa)\times 100\%$ confidence interval for the spatial dependence parameter $\eta$ is given by
\begin{align}
	\hat{\eta} \pm z_{1-\kappa/2} \left\{n \left(-\hat{l}_n''(\hat{\eta}) \right) \right\}^{-1/2}, \label{eqCIeta}
\end{align}
where $\hat{\eta}$ is the final estimate from \autoref{algorithm:est2} and
$z_{1-\kappa/2}$ is the $(1-\kappa/2)$-quantile of the standard normal distribution.

A more interesting application of the asymptotic normality in \autoref{thm:eta} might be hypothesis testing, as detecting spatial dependence in functional data is lacking in the literature.
Specifically, we can perform hypothesis testing on whether functional data exhibit spatial dependence or not, with the null hypothesis $H_0: \eta = 0$.
The test statistic is provided by
\begin{align}
	T_n = \sqrt{n} \left(-\hat{l}_n''(\hat{\eta}) \right)^{1/2} \hat{\eta}, \label{eqTSeta}
\end{align}
which asymptotically follows a standard normal distribution under the null hypothesis. 
Therefore, the null hypothesis is rejected at significance level $\kappa$ if $|T_n| > z_{1-\kappa/2}$.
This contributes to the practical utility of our method by providing a framework to test spatial dependence in functional data. 

\begin{rem}\label{remark3}
	In the hypothesis testing problem considered above, the null hypothesis is $H_0: \eta = 0$, which assesses the absence or presence of spatial dependence for the functional data as a whole.
	To the best of our knowledge, this is the first work to develop a formal hypothesis test for this global null hypothesis in the context of spatial functional data.
	Related work has largely focused either on exploratory assessment of spatial dependence or on hypothesis testing for other null, typically local dependence.
	For exploratory assessment, \cite{romano2022spatial} and \cite{khoo2023spatial} propose diagnostic measures analogous to Moran's I in univariate spatial statistics.
	However, these approaches do not develop formal hypothesis testing procedures.
	In \cite{zhang2016functional}, the spatial parameter $\rho$ is assumed to lie in $(0,1)$, thereby precluding testing the null hypothesis corresponding to the absence of spatial dependence.
	More recently, \cite{hoshino2024functional} develop a testing procedure for the spatial dependence parameter, however the null hypothesis is formulated locally at each given $t \in \ttt$.
	Namely, denoting the spatial dependence bi-variate function in their FSAR model by $\Lambda$, 
	their local null hypothesis $H_{0,t}$ at $t \in \ttt$ states that $\Lambda(u,t) = 0$ almost everywhere $u \in \ttt$. 
	This null $H_{0,t}$ is designed to assess local dependence structure and indicates there is no influence of the neighboring regions on the focal region only at $t \in \ttt$,
	which is fundamentally distinguishable from the testing of the global dependence parameter considered here.
	Notably, our null hypothesis is stronger than the null hypothesis in \cite{hoshino2024functional},
	as it requires the absence of spatial dependence uniformly over the entire functional domain. As a result, the two approaches address different inferential questions.
\end{rem}

\section{Simulation}\label{sec5}

In this section, we present a Monte Carlo simulation study to evaluate the performance of the proposed method. 
We can consider the general model in \eqref{general_model}. However, in this formulation, the spatial dependence parameter $\eta$ should lie within the interval $(1/\lambda_{\mathrm{min}}, 1/\lambda_{\mathrm{max}})$, 
where $\lambda_{\mathrm{min}}$ and $\lambda_{\mathrm{max}}$ denote the smallest and largest eigenvalues of $\bm{W}$, respectively, 
to ensure that the matrix $(\bm{I}_n-\eta \bm{W})$ is nonsingular.
In many practical settings, however, this admissible range for $\eta$ can be quite narrow.
For example, under a four-nearest neighborhood structure, the parameter space for $\eta$ may be as restrictive as $(-0.25, 0.25)$. 
Instead, we consider a nested version of the model described in \eqref{general_model}, which allows us to examine the model's behavior across a wider range of spatial dependence strengths. 
		Let
		\begin{align*}
			\eta_{ij} = \rho {w_{ij} \over w_{i+}} , \quad \Gamma_i =  {1 \over w_{i+}} \Gamma,
		\end{align*} 
		where $\rho \in [0, 1)$ is the spatial correlation parameter, $\Gamma$ is a covariance operator from $\HH$ to $\HH$, 
		$\bm{W}=[w_{ij}]_{1\leq i,j \leq n}$ denotes the neighborhood matrix defined in (\ref{def:w}), and
		$w_{i+} = \sum_{j=1}^n w_{ij}$ is the number of neighbors at location $\mbs_i$.
		This specification is analogous to the nested formulation of the univariate CAR model, which is introduced to ensure the nonsingularity of the covariance matrix \citep[cf.][]{Clayton1996, banerjee2003hierarchical, gelfand2003proper, wall2004close, ver2018spatial, shen2022conditional}.
		It leads to the following full conditional distribution:
		\begin{align}\label{model:eg2_intrinsic:full conditional}
			Y(\mbs_i) | Y(\mbs_{-i}) \sim \mathsf{Gaussian} \left(\mu_i, {1 \over w_{i+}} \Gamma \right),
		\end{align}		
		with the conditional mean function
		\begin{align}\label{model:eg2_intrinsic: mean}
			\mu_i = \alpha +   {\rho \over w_{i+}} \sum_{j \in N_i} (Y(\mbs_j)-\alpha).
		\end{align}	
	We note that this is a nested version of the model described in \eqref{general_model}, and thus the theoretical results developed for the latter also apply to the former.

Since there are no existing methods for FCAR models in a frequentist way, 
regarding the covariance estimation, 
we compare our proposed estimator from \autoref{algorithm:est2} (cf.~Equation~\eqref{eqGaHatCondCent}) with the naive one given in \eqref{eqGaHatMargCent},
which are hereafter denoted respectively as $\hat{\Gamma}_{\mathrm{FCAR}}$ and $\hat{\Gamma}_{\mathrm{naive}}$.

In each Monte Carlo simulation, random samples $\{Y(\mbs_i)\}_{i=1}^n$ were generated on a regular lattice wrapped on a torus of size $10\times 10$, $20\times 20$, $30\times 30$, and $40\times 40$, corresponding to lattice or sample sizes $n=100, 400, 900$, and $1600$.
A four-nearest neighborhood structure was employed to define spatial dependence.

The random samples were generated by a Gibbs sampling algorithm based on the full conditional distribution given in (\ref{model:eg2_intrinsic:full conditional}).
In the Gibbs algorithm, initial values were drawn from independent Gaussian distributions with mean 0 and variance 1. After a burn-in period of 200 iterations, the 200\textit{th} dataset was collected for analysis. The Gibbs algorithm was re-initialized in each Monte Carlo simulation. 

Within the Gibbs sampler, each curve was simulated using a truncated Karhunen-Lo\`{e}ve expansion:
\begin{align*}
	Y(\mbs_i)&= \mu_i + {1\over \sqrt{w_{i+}}} \sum_{j=1}^{15}  \lambda_{j}^{1/2} \xi_j \phi_{ j},
\end{align*}
where $\mu_i$ was defined in (\ref{model:eg2_intrinsic: mean}) with $\alpha= 0$. 
Similar to \cite{hall2007methodology}, we set the eigenvalues $\lambda_{j} = j^{-b/2}$ for $j=1,\cdots, 15$, with $b \in \{ 2,4 \}$, the trigonometric basis functions $\{\phi_{j}(t)\}_{j=1}^{15}$, and independent functional principal component scores $\xi_j \sim \nd(0,1)$.
All curves were produced at 50 equally spaced points over the interval $[0,1]$.

The performance  of the estimators is evaluated by finding Monte Carlo approximations of the following mean squared errors (MSEs):
\begin{align*}
	\mathrm{MSE}_{\mathrm{HS}}(\hat{\Gamma}_l)
	& = \eo[\|\hat{\Gamma}_l-\Gamma\|_{\mathrm{HS}}^2],
	\quad l \in \{\mathrm{FCAR},\mathrm{naive}\}, \\
	\mathrm{MSE}(\hat{\rho}) & = \eo[|\hat{\rho}-\rho|^2], \\
	\mathrm{MSE}(\hat{\alpha})
	& = \eo[\|\hat{\alpha}-\alpha\|_{\HH}^2].  
\end{align*}
Here, $\hat{\Gamma}_{\mathrm{FCAR}}$ and $\hat{\rho}$ are the final estimates from \autoref{algorithm:est2} (cf.~Equations~\eqref{eqGaHatCondCent} and~\eqref{eqEtaHatProjLik}) with
$\hat{\alpha} = n^{-1} \sum_{i=1}^n Y(\bm{s}_i)$,
while $\hat{\Gamma}_{\mathrm{naive}}$ is the naive estimator given in \eqref{eqGaHatMargCent}.
To assess inferential methods for $\rho$, 
we approximate the covarage probability of the interval $\widehat{\mathrm{CI}}$ in  \eqref{eqCIeta} and the rejection rate of the statistic $T_n$ in \eqref{eqTSeta}, respectively:
\begin{align*}
	\mathrm{CR}(\widehat{\mathrm{CI}}) 
	& = \pr(\rho \in \widehat{\mathrm{CI}}), \\
	\mathrm{RR}(T_n) 
	& = \pr(|T_n| \geq z_{1-\kappa/2}).
\end{align*}
In particular, the test with $T_n$ is compared with the standard approach for univariate spatial data, called Moran's I test; 
we apply Moran's I test using the average of the functional observations, $Y_{\mathrm{mean}}(\mbs_i) = {T^{-1}} \sum_{t} Y(\mbs_i)(t)$, and the $p$-value is used to similarly compute the empirical rejection rate. 
For confidence interval and hypothesis testing, we target a significance level $1-\kappa$ and a size $\kappa$, respectively, where $\kappa = 0.05$.

The following results are based on $M=500$ Monte Carlo simulations, with convergence criteria of $\tau_1=10^{-4}$ and $\tau_2=10^{-3}$.
The number of iterations required for convergence was typically between 3 and 5 across all sample sizes $n \in \{100, 400, 900, 1600\}$. Occasionally, up to 6 iterations were needed for $n=100$. 
These findings demonstrate the practical computational feasibility of our method, as evidenced by its consistently fast convergence across various sample sizes.
Furthermore, we use the FVE with a threshold of 0.95, as stated in \autoref{algorithm:est2}, to select the truncation level $p$. 
A more detailed discussion is provided in Section~S4 of the supplement.

\autoref{table:G_mise} reports the MSEs of the covariance estimators $\hat{\Gamma}_l$ for $l \in \{ \mathrm{FCAR}, \mathrm{naive} \}$. 
Most importantly, across various levels of spatial dependence, the proposed estimator $\hat{\Gamma}_{\mathrm{FCAR}}$ using conditionally centered data substantially performs better than or is comparable with the naive counterpart $\hat{\Gamma}_{\mathrm{naive}}$ based on marginally centered data,
showing its stable behavior at the same time. 
In contrast, strong dependence with large $\rho$ significantly deteriorates the performance of the naive estimator $\hat{\Gamma}_{\mathrm{naive}}$ even with large sample sizes. 
Additionally, smoothness determined by the eigendecay rate $b$ affects the performance of the covariance estimators.
In general, higher values of $b$ (i.e., faster decay) yield lower MSEs due to increased concentration on leading eigencomponents, 
which could improve accuracy.

\begin{table}[t!]
	\centering
	\caption{
		MSEs of the covariance estimators $\hat{\Gamma}_{\mathrm{FCAR}}$ and $\hat{\Gamma}_{\mathrm{naive}}$, 
		where $b \in \{2, 4\}$, $\rho\in \{0, 0.3, 0.6, 0.9, 0.99\}$, and $n\in \{100, 400, 900, 1600\}$.
	}
	\begin{tabular}{rr
			cc cc cc cc}
		\toprule
		$b$ & $\rho$ 
		& \multicolumn{2}{c}{$n=100$} 
		& \multicolumn{2}{c}{$n=400$} 
		& \multicolumn{2}{c}{$n=900$} 
		& \multicolumn{2}{c}{$n=1600$} \\
		\cmidrule(r){3-4} \cmidrule(r){5-6} \cmidrule(r){7-8} \cmidrule(){9-10}
		&  
		& $\hat{\Gamma}_{\mathrm{naive}}$& $\hat{\Gamma}_{\mathrm{FCAR}}$
		& $\hat{\Gamma}_{\mathrm{naive}}$ & $\hat{\Gamma}_{\mathrm{FCAR}}$
		& $\hat{\Gamma}_{\mathrm{naive}}$ & $\hat{\Gamma}_{\mathrm{FCAR}}$
		& $\hat{\Gamma}_{\mathrm{naive}}$ & $\hat{\Gamma}_{\mathrm{FCAR}}$\\
		\midrule
		2 & 0   & 0.127 & {0.128 }& 0.031 & {0.031} & 0.014 & {0.014} & 0.008 & {0.008} \\
		& 0.3 & 0.136 & {0.135} & 0.034 & {0.032} & 0.016 & {0.014} & 0.009 & {0.008} \\
		& 0.6 & 0.191 & {0.142} & 0.061 & {0.034} & 0.040 & {0.015} & 0.031 & {0.008} \\
		& 0.9 & 0.601 & {0.157} & 0.395 & {0.038} & 0.357 &{ 0.017} & 0.351 & {0.009} \\
		&0.99 &  1.372     &  {0.154}      &  1.849      &  {0.038}    &  1.990     &  {0.017}    & 2.052  & {0.010} \\
		\midrule
		4 & 0   & 0.037 & {0.037} & 0.009 & {0.009} & 0.004 & {0.004} & 0.002 & {0.002} \\
		& 0.3 & 0.038 & {0.036} & 0.010 & {0.009} & 0.005 & {0.004} & 0.003 & {0.002} \\
		& 0.6 & 0.063 & {0.040} & 0.026 & {0.010} & 0.020 & {0.005} & 0.017 & {0.002} \\
		& 0.9 & 0.269 & {0.046} & 0.225 & {0.010} & 0.228 & {0.005} & 0.224 & {0.003} \\
		&0.99 & 0.701      &    {0.044}   &   1.063     &   {0.011}   &   1.207    &   {0.005}   &  1.321 &  {0.003}\\
		\bottomrule
	\end{tabular}
	\label{table:G_mise}
\end{table}

Results for the spatial dependence parameter $\rho$ are presented in  \autoref{table:rho}. 
The estimator $\hat{\rho}$ is quite accurate as its MSEs are less than $1\%$ of the true parameter values across all scenarios.
The 95\% confidence interval \eqref{eqCIeta} based on $\hat{\rho}$ also exhibit strong performance.
Both the estimator and confidence interval  tend to improve as the sample size $n$ increases,
aligning well with the  theoretical results in \autoref{thm:eta}.
Slight undercoverage is observed under a few scenarios with strong spatial dependence.
This decrease in coverage may be explained by the behavior of the negative second derivative of the log-likelihood, denoted as $-\hat{l}''_n(\hat{\rho})$. 
As $\rho$ increases, the standard deviation term ${n(-\hat{l}''_n(\hat{\rho}))}^{-1/2}$, which influences the width of confidence intervals, diminishes according to the explicit form of $-\hat{l}''_n(\hat{\rho})$. 
This reduction results in narrower intervals, potentially leading to undercoverage.

\begin{table}[b!]
	\centering
	\caption{
		MSEs of the spatial dependence parameter estimator $\hat{\rho}$ and coverage rates (CR) of the $100(1-\kappa)\%$ confidence intervals given in \eqref{eqCIeta} for $\rho$, where $b \in \{2, 4\}$, $\rho\in \{0, 0.3, 0.6, 0.9, 0.99\}$, $n\in \{100, 400, 900, 1600\}$, and $\kappa = 0.05$.}
	\begin{tabular}{rr
			cc cc cc cc}
		\toprule
		$b$ & $\rho$ 
		& \multicolumn{2}{c}{$n=100$} 
		& \multicolumn{2}{c}{$n=400$} 
		& \multicolumn{2}{c}{$n=900$} 
		& \multicolumn{2}{c}{$n=1600$} \\
		\cmidrule(r){3-4} \cmidrule(r){5-6} \cmidrule(r){7-8} \cmidrule(){9-10}
		&  
		& MSE & CR 
		& MSE & CR
		& MSE & CR
		& MSE & CR \\
		\midrule
		2 & 0   & 0.004 & 0.968 & 0.001 & 0.964 & 0.000 & 0.978 & 0.000 & 0.968 \\
		& 0.3 & 0.006 & 0.916 & 0.001 & 0.940 & 0.001 & 0.954 & 0.000 & 0.952 \\
		& 0.6 & 0.004 & 0.868 & 0.001 & 0.924 & 0.000 & 0.940 & 0.000 & 0.946 \\
		& 0.9 & 0.001 & 0.882 & 0.000 & 0.910 & 0.000 & 0.938 & 0.000 & 0.922 \\
		& 0.99 &  0.000 & 0.908  & 0.000 & 0.922 & 0.000 & 0.928 & 0.000 & 0.950 \\
		\midrule
		4 & 0   & 0.006 & 0.972 & 0.001 & 0.976 & 0.001 & 0.962 & 0.000 & 0.962 \\
		& 0.3 & 0.011 & 0.930 & 0.002 & 0.956 & 0.001 & 0.952 & 0.001 & 0.950 \\
		& 0.6 & 0.007 & 0.930 & 0.002 & 0.944 & 0.001 & 0.926 & 0.000 & 0.950 \\
		& 0.9 & 0.002 & 0.912 & 0.000 & 0.936 & 0.000 & 0.942 & 0.000 & 0.954 \\
		& 0.99 &  0.000 & 0.958 & 0.000 & 0.918 & 0.000 & 0.964 & 0.000 &  0.962\\
		\bottomrule
	\end{tabular}
	\label{table:rho}
\end{table}

\begin{figure}[t]
	\centering
	\includegraphics[width=0.6\linewidth]{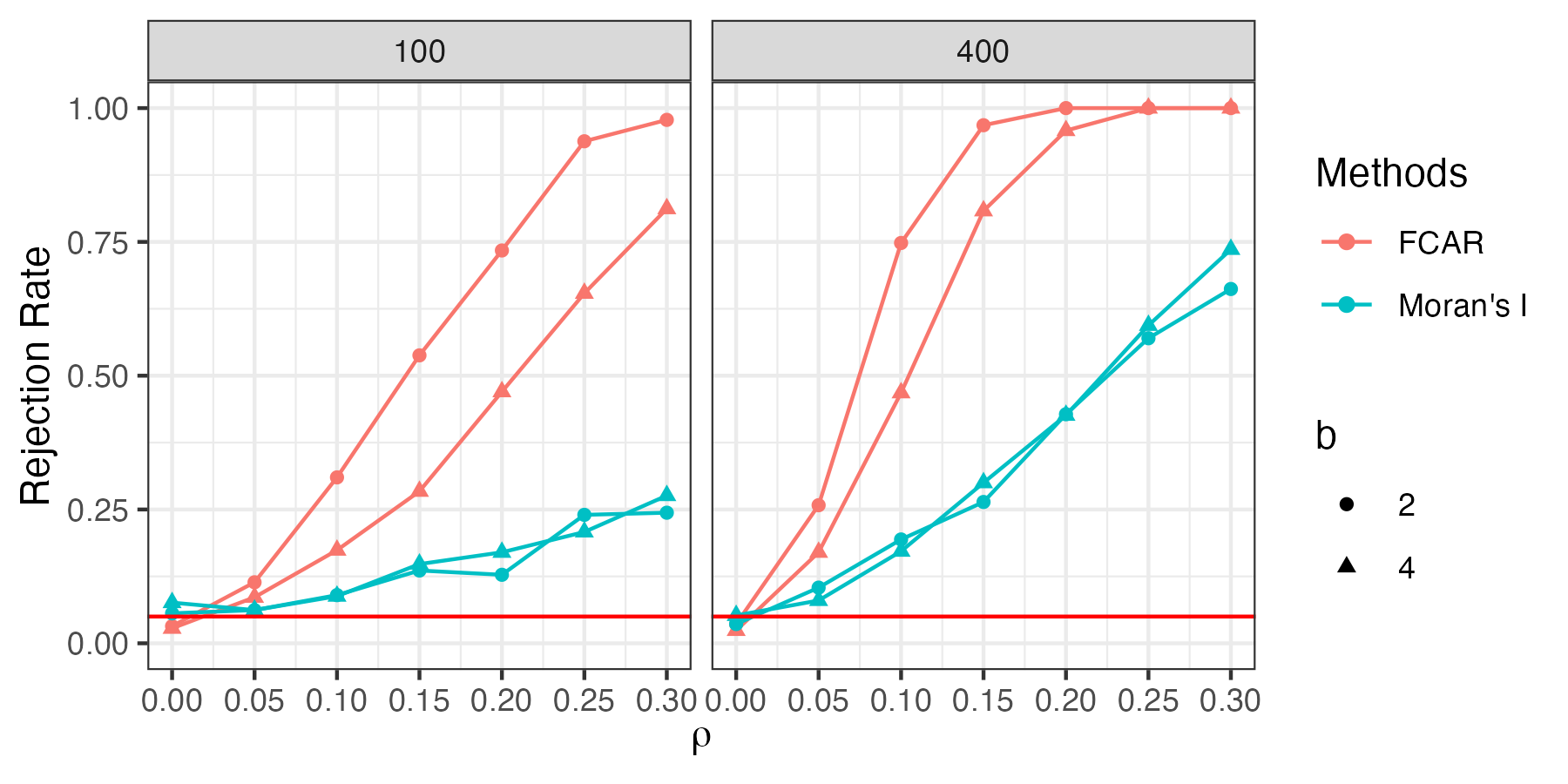}
	\caption{
		Rejection rates of the proposed test statistic \eqref{eqTSeta} based on the FCAR models and Moran's I test applied to the averaged univariate response.
		The sample sizes under consideration are $n=100$ (left) and $n=400$ (right),
		and the red horizontal lines indicate the nominal significance level of $\kappa = 0.05$.}
	\label{fig:simulation_power}
\end{figure}

\autoref{fig:simulation_power} displays the rejection rates of the two methods:
our proposed test based on the statistic \eqref{eqTSeta} and Moran’s I test applied to a univariate response
defined as the average of the functional observations over time, $Y_{\mathrm{mean}}(\mbs_i) = {T^{-1}} \sum_{t} Y(\mbs_i)(t)$. 
The proposed test consistently demonstrates higher empirical power while maintaining correct sizes,
even under challenging conditions such as weak spatial dependence (small $\rho$) and small sample sizes.
In contrast, Moran’s I test exhibits lower power across all scenarios and fails to attain the nominal significance level in some cases, particularly when $\rho = 0$.
These findings highlight the practical utility of the proposed test based on the FCAR models in detecting spatial dependence.
In general, the power of both tests tends to increase as the decay rate $b$ decreases and the sample size $n$ increases.

Finally, in \autoref{table:alpha_mise}, the estimation of $\alpha$ by $\hat{\alpha} = n^{-1} \sum_{i=1}^n Y(\bm{s}_i)$  is quite accurate in all settings, 
with decreasing MSE of $\hat{\alpha}$ as the sample size $n$ increases.

\begin{table}[b!]
	\centering
	\caption{
		MSEs of $\hat{\alpha} = n^{-1} \sum_{i=1}^n Y(\bm{s}_i)$,
		where $b \in \{2, 4\}$, $\rho\in \{0, 0.3, 0.6, 0.9, 0.99\}$, and $n\in \{100, 400, 900, 1600\}$.}
	\begin{tabular}{rrcccc}
		\toprule
		$b$ & $\rho$ & $n=100$ & $n=400$ & $n=900$ & $n=1600$ \\
		\midrule
		2 & 0   & 0.008 & 0.002 & 0.001 & 0.001 \\
		& 0.3 & 0.012 & 0.003 & 0.001 & 0.001 \\
		& 0.6 & 0.021 & 0.005 & 0.002 & 0.001 \\
		& 0.9 & 0.082 & 0.020 & 0.009 & 0.005 \\
		& 0.99 & 0.818 & 0.205 & 0.091 &  0.050\\
		\midrule
		4 & 0   & 0.004 & 0.001 & 0.000 & 0.000 \\
		& 0.3 & 0.006 & 0.001 & 0.001 & 0.000 \\
		& 0.6 & 0.010 & 0.002 & 0.001 & 0.001 \\
		& 0.9 & 0.040 & 0.011 & 0.004 & 0.003 \\
		& 0.99 & 0.370 & 0.094 & 0.043 & 0.024 \\
		\bottomrule
	\end{tabular}
	\label{table:alpha_mise}
\end{table}

\section{Real data analysis}\label{sec6}
\begin{figure}[b!]
	\centering
		\begin{minipage}[t]{0.48\textwidth}
			\includegraphics[width=\textwidth]{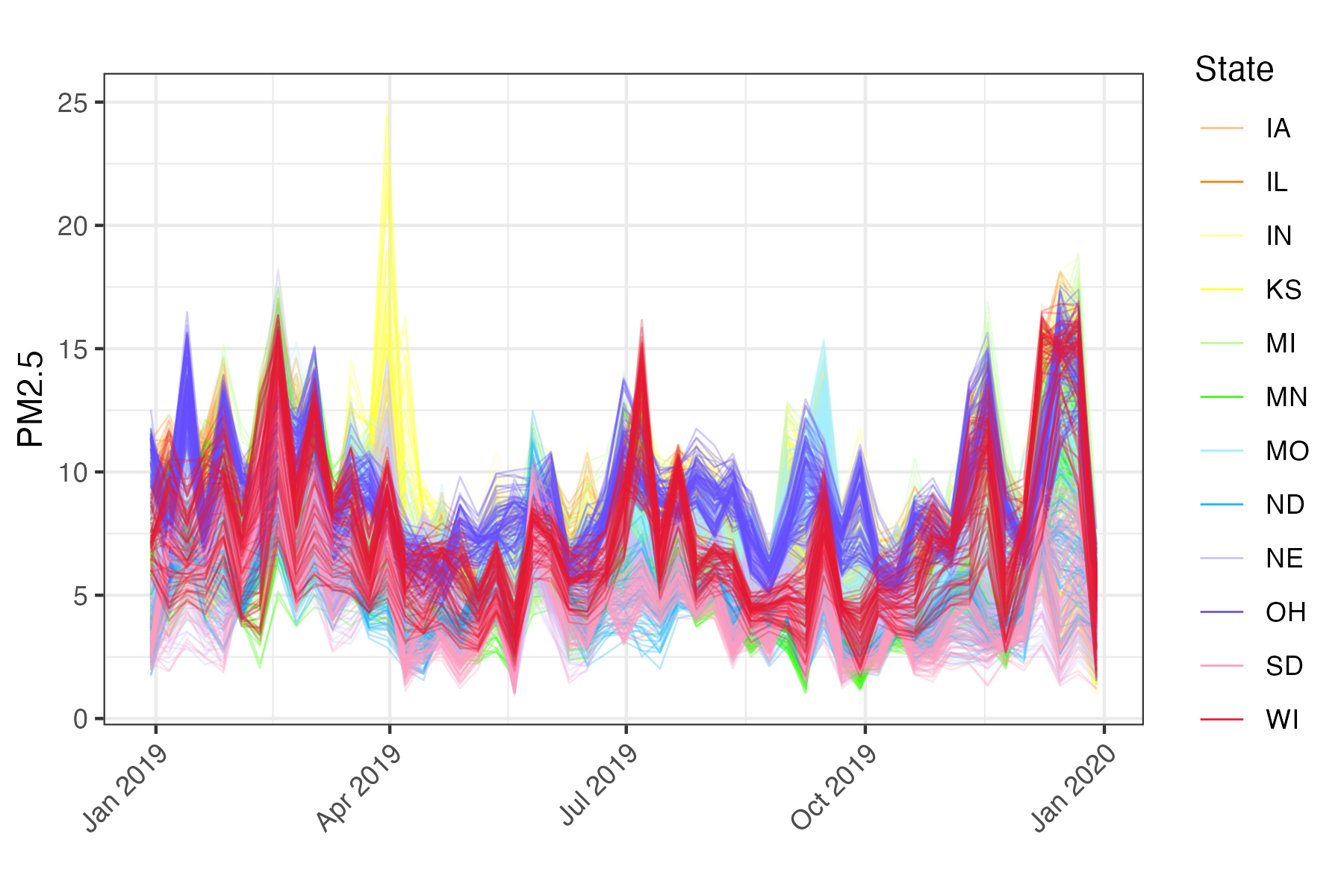} 
			\centering
			\text{(a)}
		\end{minipage}
		\hfill
		\begin{minipage}[t]{0.48\textwidth}
			\includegraphics[width=\textwidth]{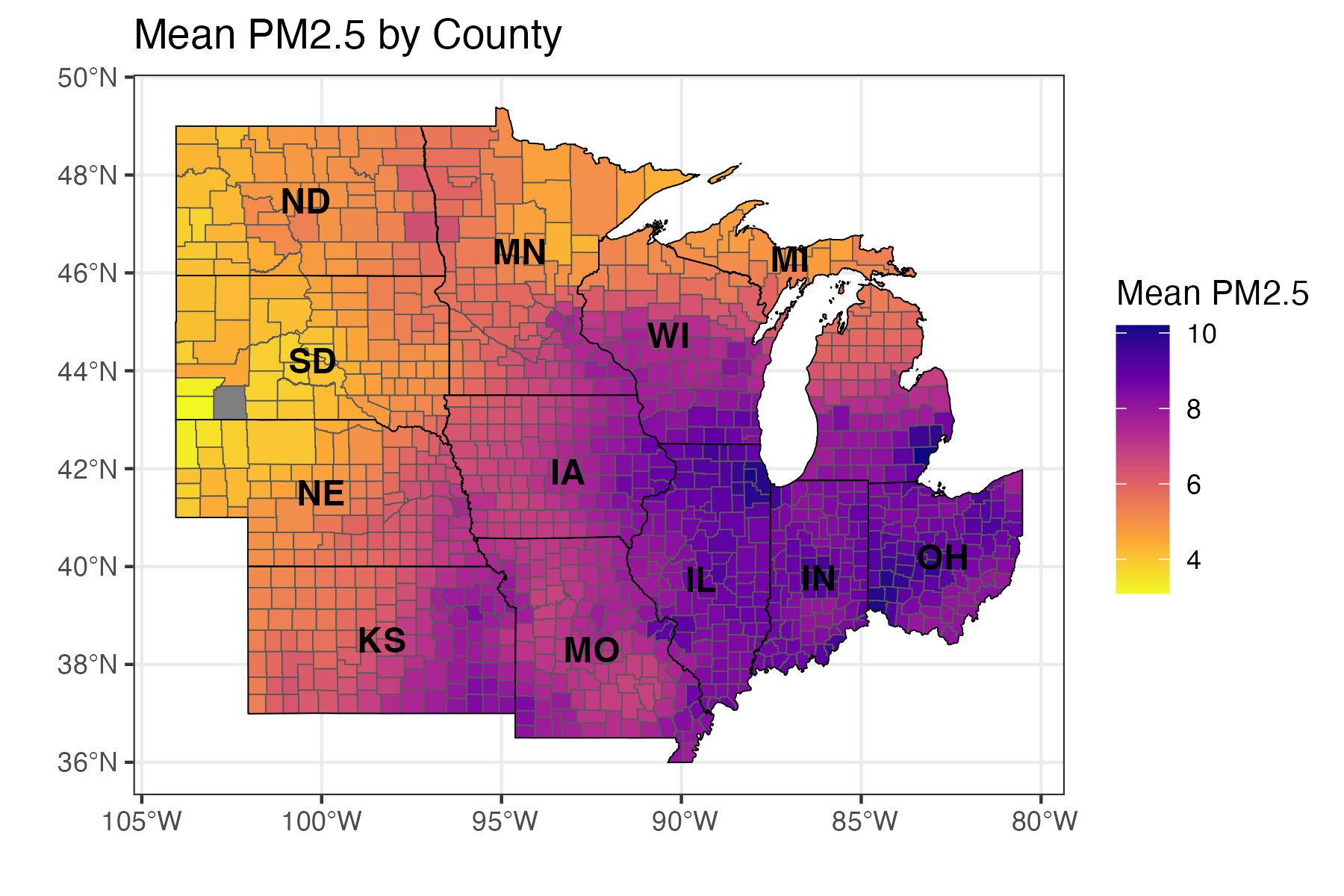} 
			\centering
			\text{(b)}
		\end{minipage}
	\caption{
		PM$_{2.5}$ concentrations in 1054 counties in the Midwestern United States.
		(a) Weekly curves over 53 weeks;
		(b) Average concentration for 53 weeks.}
	\label{fig:data_pm_plot}
\end{figure}

We apply the proposed methods based on our new FCAR models to a real dataset of PM$_{2.5}$ concentrations in the Midwestern United States, obtained from the Centers for Disease Control and Prevention (CDC)\footnote{https://data.cdc.gov/Environmental-Health-Toxicology/Daily-County-Level-PM2-5-Concentrations-2001-2019/dqwm-pbi7/about\_data}.
The functional data was defined by the weekly average PM$_{2.5}$ values recorded over 53 weeks, from December 30, 2018, to December 29, 2019,  across 1054 counties in the study region. \autoref{fig:data_pm_plot}~(a) presents the PM$_{2.5}$ concentration curves, with lines colored by state. 
We observe that counties within the same state tend to exhibit similar temporal patterns.
To apply our proposed FCAR method based on the model described in \eqref{model:eg2_intrinsic:full conditional}, the adjacent region was defined as neighbors.
For illustration, the average PM$_{2.5}$ concentration over 53 weeks is depicted in \autoref{fig:data_pm_plot}~(b). 
Notably, the elevated averaged PM$_{2.5}$ values are observed in major cities such as Chicago in Illinois, Indianapolis in Indiana, Cincinnati in Ohio, and Detroit in Michigan.

\begin{figure}[t]
	\centering
	\includegraphics[width=0.7\linewidth]{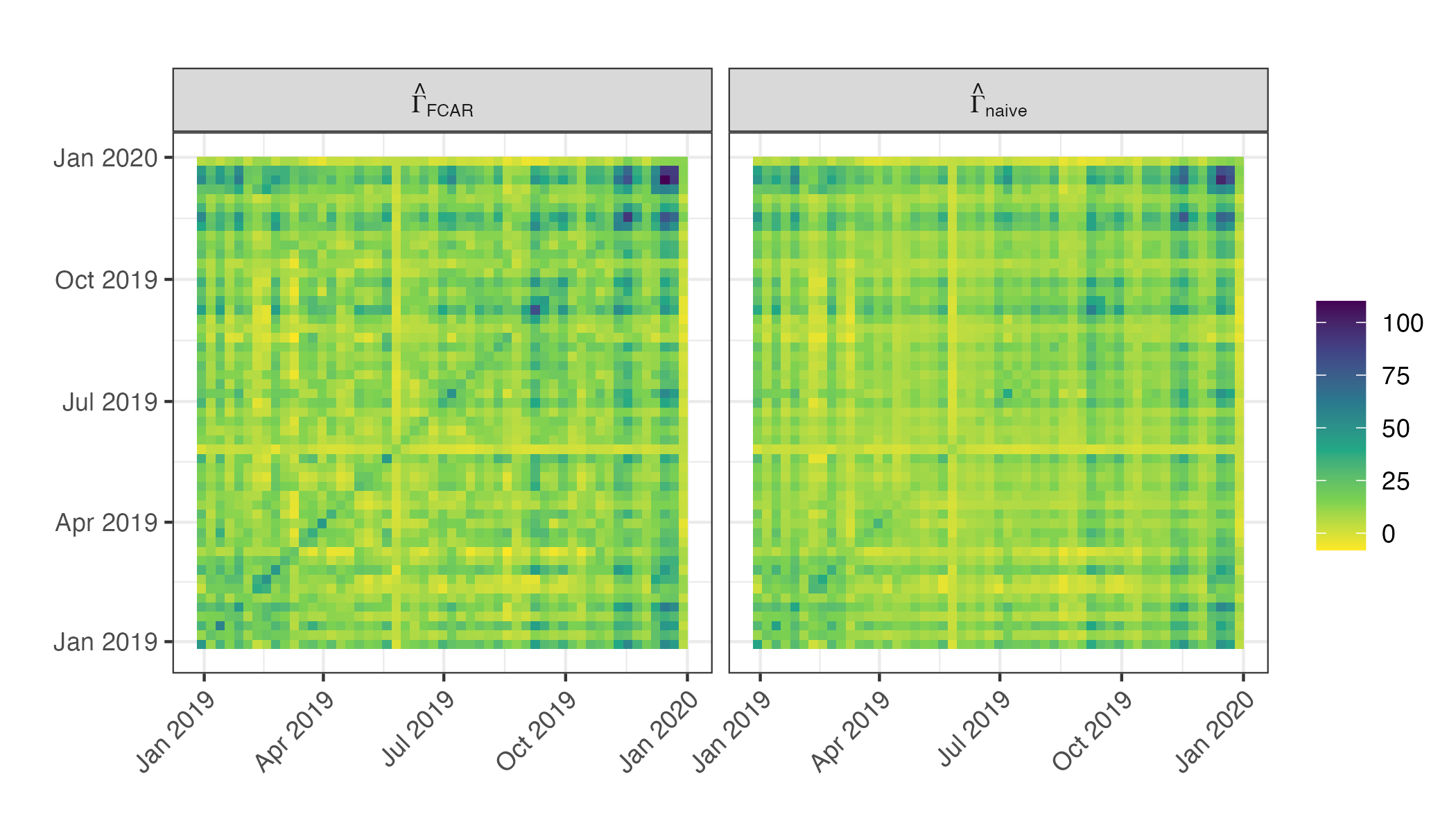}
	\caption{Heatmaps of estimated covariance operators from PM$_{2.5}$ data in the Midwestern United States: $\hat{\Gamma}_{\mathrm{FCAR}}$ (left) and $\hat{\Gamma}_{\mathrm{naive}}$ (right).}
	\label{fig:data_GHat_plot}
\end{figure}

\begin{figure}[b!]
	\centering
	\includegraphics[width=0.5\linewidth]{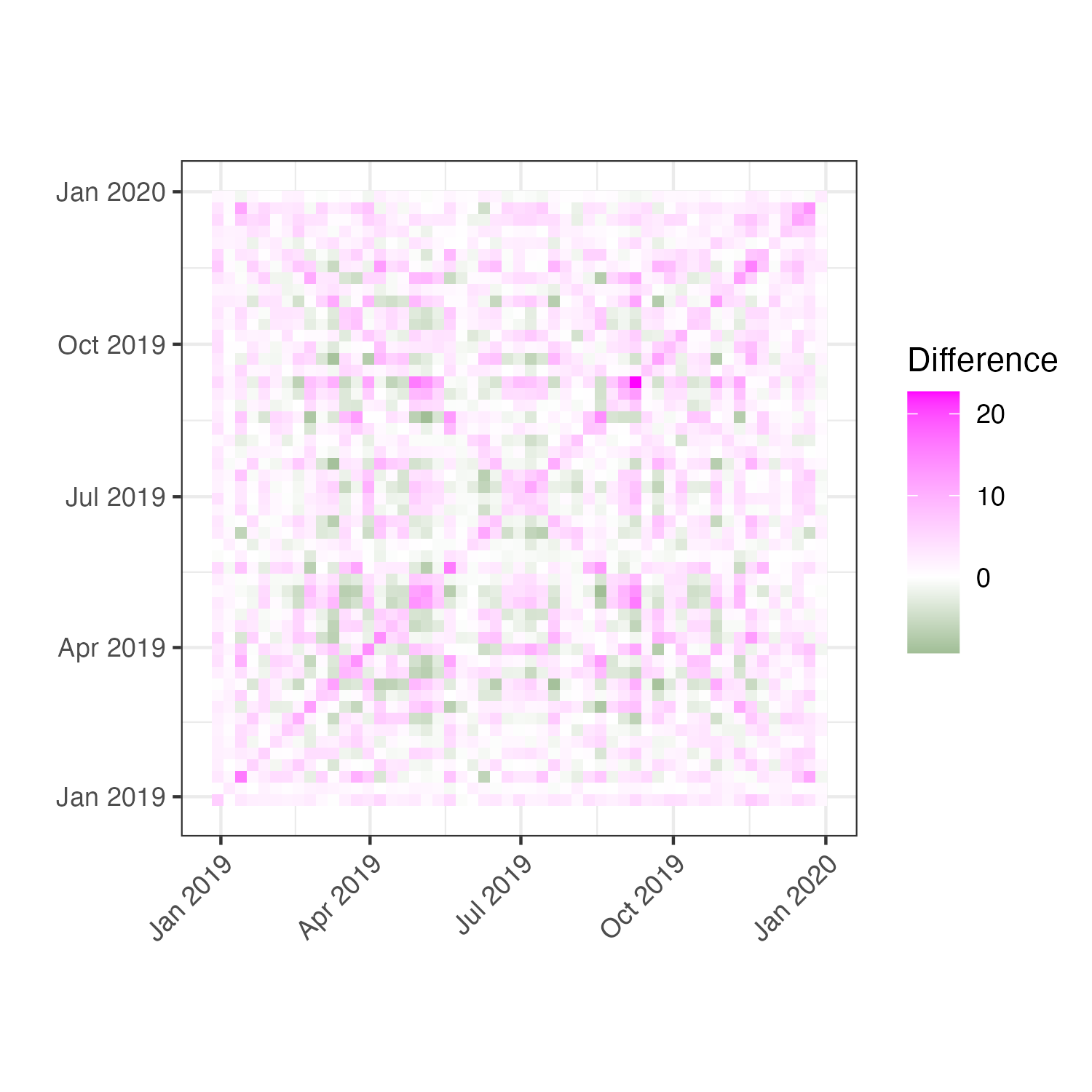}
	\caption{Heatmap of the element-wise differences between the estimated covariance operators $\hat{\Gamma}_{\mathrm{FCAR}}$ and $\hat{\Gamma}_{\mathrm{naive}}$, i.e., $\hat{\Gamma}_{\mathrm{FCAR}}-\hat{\Gamma}_{\mathrm{naive}}$  from PM$_{2.5}$ data in the Midwestern United States.}
	\label{fig:data_GHat_diff_plot}
\end{figure}

The estimated covariance operators are compared in \autoref{fig:data_GHat_plot}:
the final covariance estimator  $\hat{\Gamma}_{\mathrm{FCAR}}$ from \autoref{algorithm:est2} and
the naive estimator $\hat{\Gamma}_{\mathrm{naive}}$ from \eqref{eqGaHatMargCent}.
It reveals that the estimated covariance operator $\hat{\Gamma}_{\mathrm{FCAR}}$ exhibits greater structural complexity, with stronger values concentrated near the diagonal, 
which indicate pronounced short-range temporal correlations.
On the other hand, the naive covariance estimator $\hat{\Gamma}_{\mathrm{naive}}$ shows a simpler and more diffuse pattern.
In \autoref{fig:data_GHat_diff_plot}, 
the element-wise differences between the two estimates, $\hat{\Gamma}_{\mathrm{FCAR}}-\hat{\Gamma}_{\mathrm{naive}}$, are displayed.
In December 2019---when the PM$_{2.5}$ concentrations exhibited very high variability in \autoref{fig:data_pm_plot}~(a)---both models captured this variation well, leading to the smallest differences among the diagonals.
However, in September 2019, although the observed variability in PM$_{2.5}$ levels was also relatively high, it was not as much as extreme as in December 2019. 
The FCAR method estimated a relatively high variance of 81.69 at this time point, while the naive one yielded a substantially lower estimate of 58.98. It leads to the largest discrepancy in \autoref{fig:data_GHat_diff_plot}.
Since the estimated spatial dependence is extremely high, 
we conclude that the naive method underestimates low to moderate variation across counties by ignoring the spatial dependence.

The estimated spatial dependence parameter was $\hat{\rho}=0.999$, with a 95\% confidence interval of $(0.999, 1)$.
Also, the hypothesis test with null $\rho=0$ yielded a $p$-value of 0, which implies strong evidence of spatial dependence.
On the other hand, the Moran's I statistic and the corresponding $p$-value based on the average PM$_{2.5}$ were 0.035 and 0.023, respectively.
At a 5\% nominal significance level, both the FCAR method and Moran’s I test reject the null hypothesis of no spatial dependence. 
However, at a more stringent 1\% level, only the FCAR method identifies significant spatial structure, whereas Moran’s I test fails to reject the null. 
This suggests that incorporating the full functional trajectory can reveal spatial dependence patterns that may not be detected by Moran’s I applied to temporally averaged responses at the same significance level.

\begin{figure}[t]
	\centering
	\includegraphics[width=0.5\linewidth]{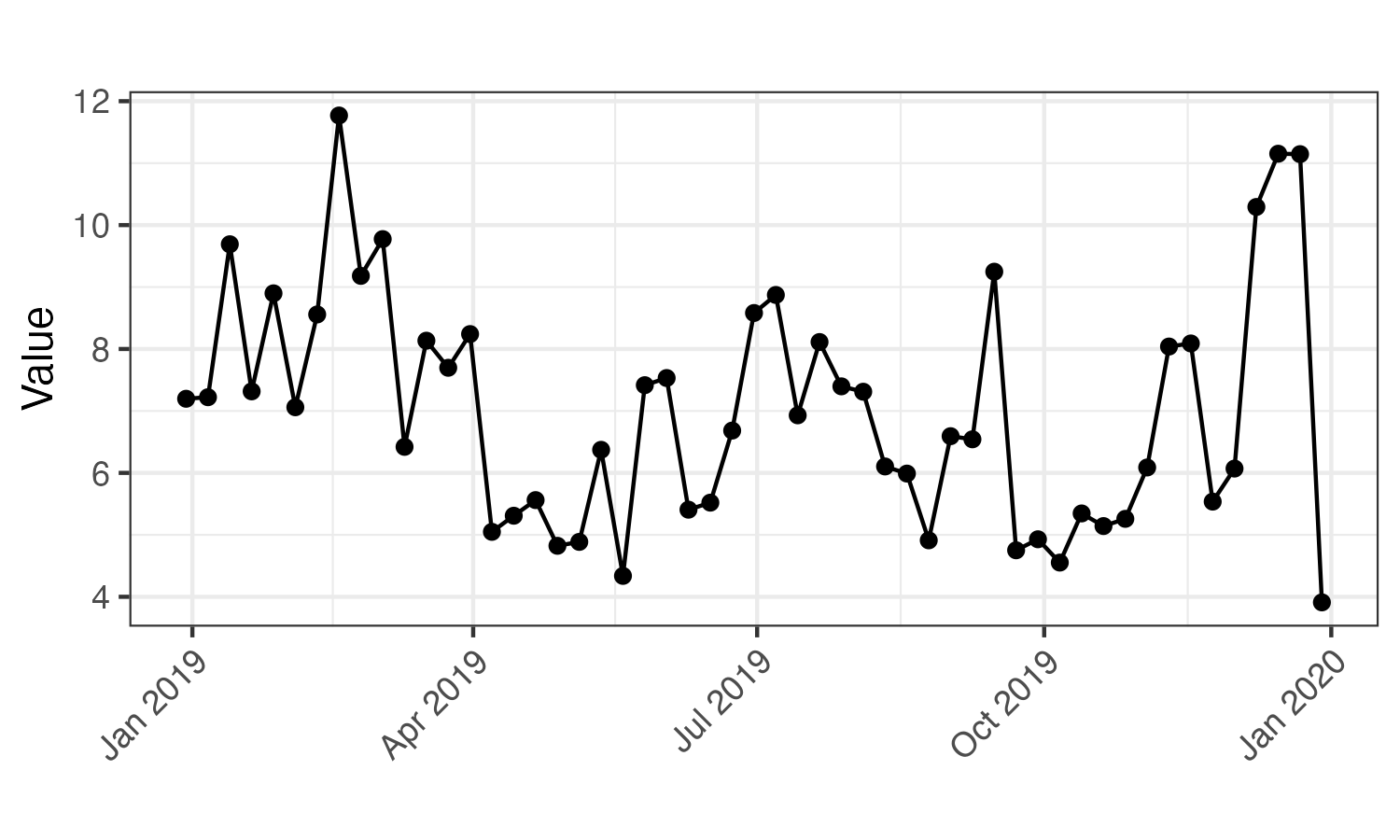}
	\caption{Estimated parameter function $\hat{\alpha}$ from PM$_{2.5}$ data in the Midwestern United States.}
	\label{fig:data_alphaHat_plot}
\end{figure}

\autoref{fig:data_alphaHat_plot} illustrates the estimated parameter function $\hat{\alpha}(t)$, which peaks during the winter months and reaches lower values in the summer.
This seasonal pattern may be attributed to increased heating and fuel consumption in colder months, while enhanced air circulation, precipitation, and reduced heating activity contribute to lower values during the summer.


\section{Conclusion}\label{sec7}


In spatial functional data analysis, two widely used classes of models are FCAR and functional simultaneous autoregressive (FSAR) models.
FSAR models have primarily been developed in the context of scalar-on-function regression, 
where the predictor is a function and the response is scalar \citep[cf.][]{pineda2019functional, ahmed2022quasi}.
Recently, \cite{hoshino2024functional} introduced FSAR model with functional responses.
In that work, the FSAR model is formulated using a spatial dependence operator rather than a scalar spatial dependence parameter, 
leading to a hypothesis testing framework that is conceptually different from the one considered here (cf. \autoref{remark3}).
On the other hand, existing FCAR models remain limited. 
The FCAR model of \cite{zhang2016functional} relies on basis expansion and truncation strategy at the modeling stage.
By extending the conditional model specification of univariate CAR models directly into the functional data framework,
our proposed FCAR models avoids potential model bias associated with modeling-stage truncation.

While we consider the binary neighborhood structure in \eqref{def:w}, the theoretical framework is not tied to binary weights per se and may be extended to more general spatial weight matrices, such as distance-based weights \citep[cf.][Equation (6.5.9)]{cressie2015statistics}. 
In such cases, Assumption~\ref{A1}, which currently requires a uniformly bounded number of neighbors, should be replaced by an analogous bounded-weight condition. 
For example, for a general spatial weight matrix $\bm{W} = [w_{ij}]_{1 \leq i, j \leq n}$, one may impose
$\limsup_n \max_{1\leq i \leq n} \sum_{j \in N_i} |w_{ij}| < \infty$.
Assumption~\ref{A4}, on the other hand, already applies to a general weight matrix $\bm{W}$, 
since it requires the CAR covariance matrix $\|(I_n - \eta \bm{W})^{-1}\|_{\mathrm{F}}$ to remain asymptotically bounded. 
Thus, distance-based weights can be incorporated provided that the resulting weight matrix satisfies suitable boundedness and covariance-stability conditions. 
For instance, inverse-distance weights may need to be normalized, truncated to local neighborhoods, or chosen with sufficient decay so that the weighted row sums remain uniformly bounded under the expanding lattice asymptotic framework.

With regard to computation, the proposed method remains practically feasible even for large spatial functional datasets.
In simulations with functional observations evaluated at 365 time points across $n=900$ locations, the computation time was approximately 1.57 seconds per dataset. 
When the sample size increased to $n=1600$ locations on a $40\times 40$ regular lattice and to $n=2500$ locations on a $50\times 50$ regular lattice, 
the corresponding computation times increased to approximately 3.9 seconds and 11.5 seconds, respectively. 
Computation times were obtained using a single core of an Apple M3 Pro (11-core CPU, 18 GB RAM).
These results indicate that the proposed method is practically feasible for datasets comprising several thousand spatial locations and densely observed functional trajectories, such as daily measurements over a year.
To support reproducibility, the R package \texttt{FCARfreq} is publicly available at: \href{https://github.com/srkim3487/FCARfreq}{https://github.com/srkim3487/FCARfreq}.

Several extensions of interest can be considered in future work.
One direction would be to accommodate more complex forms of spatial dependence.
For example, the scalar spatial dependence parameter considered here could be extended to a spatial dependence operator, as in \cite{hoshino2024functional}. 
Another promising extension is to integrate the proposed FCAR framework with function-on-function regression models, 
allowing both predictors and responses to be spatially indexed functional data \citep[cf.][]{beyaztas2025spatial}.
Such extensions would further broaden the applicability of the proposed methodology to more complex spatial functional data settings.
Extensions to non-Gaussian functional spatial models are also of interest. 
One possible direction is to formulate suitable Hilbert-valued conditional distributions, for example through an exponential-family-type conditional specification, and to develop the corresponding pseudo-likelihood estimation theory. 
Unlike the Gaussian FCAR model considered in this paper, where the conditional specification yields a tractable joint distribution and supports likelihood-based estimation from projected functional observations, non-Gaussian conditional specifications require separate treatment of compatibility, joint distribution, and computational tractability;
see \cite{kim2024generalized} for analogous issues in scalar non-Gaussian spatial models.
Developing such non-Gaussian functional spatial models would be an interesting direction for future research.

\section*{Supplementary material}

\noindent
\textbf{Supplement to ``A new class of functional conditional autoregressive models":} 
This supplement provides detailed proofs of all propositions, theorems, and lemmas presented in the main manuscript. 
It also includes additional simulation results and comparisons with existing functional CAR models introduced by \cite{zhang2016functional}.

\bibliographystyle{apalike}
\bibliography{FCAR_reference}

\end{document}